\documentclass[aps,pra,twocolumn,amssymb,amsfonts,floatfix,showpacs]{revtex4}
\usepackage{bm}
\usepackage{dcolumn}
\usepackage{bm}
\usepackage{mathbbol}
\usepackage{graphicx}
\usepackage{epstopdf} 
\usepackage{amsmath,amsthm} 
\usepackage{times}
\usepackage{color}

\begin{document}

\title{Cooperative shielding in many-body systems with long-range interaction}

\author{Lea F. Santos}
\affiliation{Department of Physics, Yeshiva University, New York, New York 10016, USA \\
ITAMP, Harvard-Smithsonian Center for Astrophysics, Cambridge, MA 02138, USA}
\author{Fausto Borgonovi}
\affiliation{Dipartimento di Matematica e Fisica and ILAMP, 
Universit\'a Cattolica del Sacro Cuore, Brescia, ITALY}
\affiliation{Istituto Nazionale di Fisica Nucleare, sez. Pavia, Pavia, ITALY}
\author{Giuseppe Luca Celardo}
\affiliation{Dipartimento di Matematica e Fisica and ILAMP, 
Universit\'a Cattolica del Sacro Cuore, Brescia, ITALY}
\affiliation{Istituto Nazionale di Fisica Nucleare, sez. Pavia, Pavia, ITALY}

\date{\today}

\begin{abstract}
In recent experiments with ion traps, long-range interactions were associated with the exceptionally fast propagation of perturbation, while in some theoretical works they have also been related with the suppression of propagation. Here, we show that such apparently contradictory behavior is caused by a general property of long-range interacting systems, which we name  {\it Cooperative Shielding}. It refers to shielded subspaces that emerge as the system size increases and inside of which the evolution is unaffected by  long-range interactions for a long time. As a result, the dynamics strongly depends on the initial state:  if it belongs to a shielded subspace, the spreading of perturbation satisfies the Lieb-Robinson bound and may even be suppressed, while for initial states with components in various subspaces, the propagation may be quasi-instantaneous. We establish an analogy between the shielding effect and the onset of quantum Zeno subspaces. The derived effective Zeno Hamiltonian successfully describes the short-ranged dynamics inside the subspaces up to a time scale that increases with system size. {\it Cooperative Shielding} can be tested in current experiments with trapped ions.
\end{abstract}  

        
\maketitle


{\em Introduction.--} 
A better understanding of the nonequilibrium dynamics of many-body quantum systems is central to a wide range  of fields, from atomic, molecular, and condensed matter physics to quantum information and cosmology. New insights into the subject have been obtained thanks to the remarkable level of controllability and isolation of experiments with optical lattices~\cite{Trotzky2008,Trotzky2012,Gring2012,Fukuhara2013,Hazzard2014,Langen2015,BauerARXIV} and trapped ions~\cite{Jurcevi2014,Richerme2014}. Recently there has been a surge of interest in the dynamics of systems with long-range interactions,  triggered by experiments with ion traps~\cite{Jurcevi2014,Richerme2014}, where the range of interactions in one-dimensional (1D) spin models  can be tuned with great accuracy. Other realistic systems that contain long-range interaction include cold atomic clouds~\cite{Akkermans2008}, natural light-harvesting complexes~\cite{Grad1988,Celardo2012,Celardo2014}, helium Rydberg atoms~\cite{extra1}, and  cold Rydberg gases~\cite{extra2}.
Long-range interacting systems display features that are not often observed in other systems, such as broken ergodicity~\cite{Mukamel2005,Borgonovi2005,BorgonoviAll,Celardo2006} and long-lasting out-of-equilibrium regimes~\cite{Bachelard2008}. 

According to the usual definition~\cite{RuffoBook}, in $d$ dimension,  an interaction decaying as $1/r^{\alpha}$ (where $r$ is the distance between two bodies), is  short range  when $\alpha >d$ and it is long-range when $\alpha \le d$. 
A major topic of investigation has been whether the propagation of excitations in systems with long-range interaction remains or not confined to an effective light cone~\cite{Hastings2006,Schachenmayer2013,Eisert2013,Hauke2013,Gong2014,Mazza2014,Metivier2014,Feig2015,Storch2015}, as defined by the Lieb-Robinson bound~\cite{Lieb1972} and its generalizations~(\cite{Storch2015} and references therein).
In the aforementioned experiments with trapped ions, it was observed that for short-range interaction, the propagation of perturbation is characterized by a constant maximal velocity, being bounded to an effective light cone. As $\alpha$ decreases, the propagation velocity increases and eventually diverges. For long-range interaction, $\alpha \leq 1$, the light-cone picture is no longer valid and the dynamics becomes nonlocal. However, examples of constraint dynamics in long-range interacting systems have also been reported, including logarithmic growth of entanglement~\cite{Schachenmayer2013}, light-cone features~\cite{Storch2015}, self-trapping~\cite{Nazareno1999}, and slow decays at critical points~\cite{SantosBernal2015}.

Here, we show that these contradictory results are due to a general effect present in long-range interacting systems, which we name {\it Cooperative Shielding}. It corresponds to the onset of approximate superselection rules that cause a strong dependence of the dynamics on the initial state. Inside a superselection subspace, long-range interactions do not affect the system evolution ({\it shielding}) up to a time scale that grows with system size ({\it cooperativity}). The dynamics can then be described by an effective short-ranged Hamiltonian that either leads to a propagation within the Lieb-Robinson light cone or to localization. In contrast, for an initial state with components over several subspaces, the propagation of excitations is affected by long-range interactions and can be unbounded.

To explain how shielding can arise in a very trivial case, let us consider the total Hamiltonian $H=H_0+V$, describing a many-body quantum system, where $H_0$ has one-body terms and possible short-range interactions, and $V$ corresponds to some additional interactions. If $[H_0,V]=0$ and $V$ is highly degenerate in one of
its eigensubspaces ${\cal V}$, so that $V|V_k\rangle=v |V_k\rangle$ $\forall |v_k\rangle \in {\cal V}$, the evolution 
of any initial state $|\psi_0\rangle$ belonging to such eigensubspace is simply given by: $|\psi(t)\rangle= e^{-ivt/\hbar} e^{-iH_0t/\hbar} |\psi_0\rangle$. Since the only effect of $V$ is to induce a global phase, the dynamics  is shielded from $V$ and determined only by $H_0$. In contrast, if the initial state has large components in more than one eigensubspace of $V$, the dynamics will not be shielded from $V$. The question that we now pose is whether shielding is still possible when $[H_0,V]\neq 0$ and $V$ is no longer degenerate. We show that the answer is positive when $V$ involves only long-range interactions. The dynamics can remain shielded, but now for a finite time that increases with system size.

One can also draw a parallel between the picture above and the quantum Zeno effect (QZE). In the QZE, the dynamics of the system remains confined to subspaces tailored by the interaction with a measuring apparatus~\cite{Peres1980,Facchi2001,Facchi2001b,Facchi2002,FacchiProceed}. The stronger the interaction is, the better defined the subspaces become. Here, instead, the interaction strength is kept fixed, but due to its long-range-nature, invariant subspaces are generated. The dynamics, restricted to the  invariant subspaces, is described by a short-ranged Zeno Hamiltonian up to a time scale that diverges with system size. 


\vskip 0.2 cm
{\em The Model.--} 
We consider a 1D spin-$1/2$ model with $L$ sites and open boundary conditions described by the Hamiltonian,
\begin{eqnarray}
&&H=  H_0 + V, 
\label{ham}
\\
&&H_0 =  \sum_{n=1}^L ({\cal B} + h_n)  \sigma_n^z  
+  \sum_{n=1}^{L-1} J_z \sigma_n^z  \sigma_{n+1}^z,  \nonumber \\
&&V= \sum_{n<m}\frac{J}{|n-m|^{\alpha}} \sigma_n^x  \sigma_m^x  .\nonumber
\end{eqnarray}
Above, $\hbar=1$ and $\sigma_n^{x,y,z}$ are the Pauli matrices on site $n$.  The transverse field has a constant component ${\cal B}$ and a random part given by $h_n$, where $h_n \in [-W/2,W/2]$  are random numbers from a uniform distribution. The nearest-neighbor (NN) interaction in the $z$-direction, of strength $J_z\geq 0$, may or not be present. $J$ is the strength of the interaction in the $x$-direction with $\alpha$ determining the range of the coupling. Unless specified otherwise, $J=1$. The Hamiltonian with $W=0$ and $J_z=0$ describes the systems studied with ion traps~\cite{Jurcevi2014,Richerme2014}. In agreement with those experiments, where a limited range of system sizes is explored, $V$ is not rescaled by $L$.

When $\alpha=0$, $H$ can be written in terms of the total $x$-magnetization, $M_x = \sum_{n=1}^L \sigma_n^x/2$, as
\begin{equation}
H=  \sum_{n=1}^L ({\cal B} + h_n)  \sigma_n^z  
+  \sum_{n=1}^{L-1} J_z \sigma_n^z  \sigma_{n+1}^z 
+2JM_x^2 - \frac{JL}{2}.
\label{hamMx} 
\end{equation}
The spectrum of $V$ is divided into energy bands, each one associated with a value of the collective quantity $M_x^2$. Each band, with energy $E_b = 2J(L/2-b)^2 - JL/2$, has states with $b$ and $L-b$ excitations, where $b=0,1,\ldots L/2$. For instance, $b=1$ corresponds to states with one spin pointing up in the $x$-direction in a background of down-spins or vice-versa. An energy band contains $2\binom{L}{b}$ degenerate states if $b<L/2$ and $\binom{L}{b}$ states when $b=L/2$. In contrast,  for $0 < \alpha < 1$, the states in each band $V$ are not all degenerate anymore. 

{\em Light Cones.--} 
In Refs.~\cite{Jurcevi2014,Richerme2014}, the acceleration of the spreading of excitations and eventual surpassing of the Lieb-Robison bound achieved by decreasing $\alpha$ was verified for initial states corresponding to eigenstates of $H_0$, where each site had a spin either pointing up or down in the $z$-direction. These initial states have components in all subspaces of $V$. 

\begin{figure}[ht!]
\centering
\includegraphics*[width=3.5in]{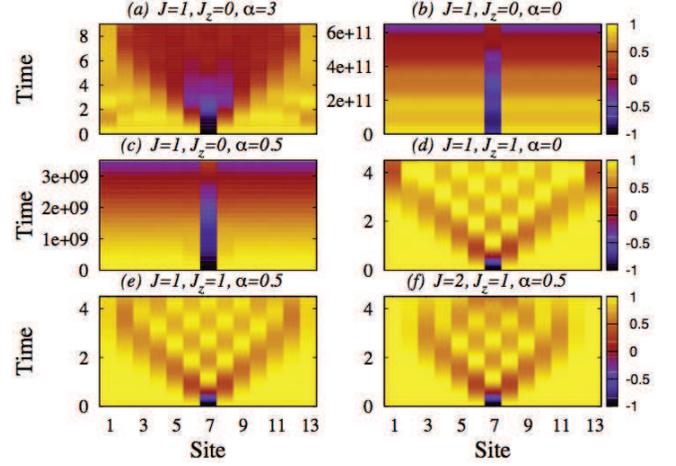} 
\caption{(Color online) Density plots for the evolution of $\langle \sigma_n^x (t) \rangle$; $L=13$; ${\cal B}=1/2$; $W=0$. Initial state: $\langle \sigma_7^x (0)\rangle =-1 $ and $\langle \sigma_{n \neq7}^x (0) \rangle = + 1$. A light cone typical of short-range interaction is seen in (a), as expected, but also in (d), (e), and (f) where the evolution is shielded from the present long-range interaction. Freezing occurs for very long times in (b); it also happens in (c) where the bands of $V$ are not degenerate.}
\label{fig:cone}
\end{figure}

Motivated by the special role of the $x$-direction in Eq.~(\ref{hamMx}) and to show the main features of {\it Cooperative Shielding}, here we change the focus of attention to initial states with spins aligned along the $x$-axis. They are the eigenstates of $V$ and are denoted by $|V_k\rangle$. In Fig.~\ref{fig:cone}, we  show the evolution of the spin polarization, $\langle \sigma_n^x (t) \rangle$,  for an initial state where all spins point up in $x$, except for the spin in the middle of the chain, which points down, so $M_x= L/2 - 1$ and $b=1$. 
 
In Fig.~\ref{fig:cone} (a), where the interaction is short range  ($\alpha=3$), $H_0$ effectively couples states belonging to different subspaces of $V$. The effects of both $H_0$ and $V$ lead to the evident light cone. This is no longer the case for long-range interaction ($\alpha<1$), as exemplified in Figs.~\ref{fig:cone} (b) and (c) for $\alpha=0$ and $0.5$. Their dynamics is frozen for a long time, which increases with the range of the interaction [compare the time scales in (b) and (c)] and with the system size (see discussion below). The long-time localization of spin excitations in Figs.~\ref{fig:cone} (b) and (c) is caused by both combined factors: the separated energy bands of $V$ and the absence of direct coupling within the band ($H_0$ is not effective and $J_z=0$). Notice that the energy bands for case (c) are no longer degenerate, yet localization persists for a long time.

Since the initial state is not an eigenstate of the total Hamiltonian, the spin excitation does eventually spread and the spins reverse their signs (see Figs.~\ref{fig:cone} (b,c) and discussion in~\cite{noteSUPPL}). This magnetic reversal can be explained in terms of macroscopic quantum tunneling~\cite{Borgonovi2005}.

While for $\alpha<1$ in the presence of an external field the dynamics is frozen, the addition of NN interaction ($J_z \neq 0$) restores the propagation of perturbations [Figs.~\ref{fig:cone} (d), (e), (f)]. Despite the existence of long-range interactions, the evolution can be described by an effective short-ranged Hamiltonian, as we show below.  This is the hallmark of the {\em Cooperative Shielding} effect discussed in this work, the suppression of propagation [Figs.~\ref{fig:cone} (b), (c)] being only a special case of it. 

In Figs.~\ref{fig:cone} (d), (e), (f), a light cone typical of short-range interactions emerges: the dynamics is independent of system size and of the long-range coupling $J$. In Fig.~\ref{fig:cone} (f),  $J$ is twice as large as in Figs.~\ref{fig:cone} (d,e), but the results in the three panels are very similar, apart from border effects. The propagation of excitations depends only on $J_z$ up to long times. This shielded evolution occurs for any $\alpha<1$ (see more figures in~\cite{noteSUPPL}). 
In the case of $0<\alpha<1$, as in Figs.~\ref{fig:cone} (e) and (f), the bands of $V$ are no longer degenerate, so the various eigenstates of $V$ that are excited within the band have different eigenenergies. One could then expect $V$ to affect the evolution, yet the velocity of propagation remains independent of $V$ for long times. This shows that the cause for shielding is not only the suppression of the transitions between different bands of $V$, but also the narrow distribution of the energies of $V$ inside the band. The motion remains constrained to subspaces that are quasi-degenerate {\em w.r.t.} to $V$. The emergence of quasi-constants of motion is recurrent in long-range interacting systems~\cite{Bachelard2008}.

{\em Invariant Subspaces and Zeno effect.--} 
Stimulated by the results of Fig.~\ref{fig:cone}, we now analyze in more details the effects of infinite-range interaction ($\alpha=0$) and their dependence on system size. 
For a general treatment, we assume a random transverse field, so ${\cal B}=0$ and $h_n\neq0$. We take as initial state $|\Psi(0)\rangle$ a random superposition of all states $|V_k^b\rangle$ that belong to the same fixed band $b$ chosen for the analysis. We verified that the results for single states $|V_k^b\rangle$ picked at random from the same energy band are equivalent.  

\begin{figure}[htb]
\centering
\includegraphics*[width=3.2in]{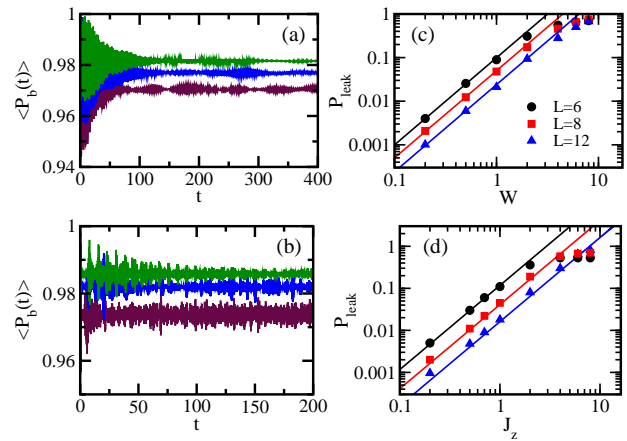}
\caption{(Color online) Probability for the initial state to remain in (a,b) or leave (c,d) its original energy band. In (a,b): $\langle P_b(t) \rangle$ for the initial random superposition of states $| V_k^b \rangle$ from band $b=1$ for $L=10,12,14$ from bottom to top;  in (a): $J_z=0,W=2$ and in (b): $J_z=1,W=0$. In (c,d): $P_{leak}$ {\em vs} $W$ for $J_z=0$ (c), and {\em vs} $J_z$ for $W=0$ (d). Symbols represent numerical results and full lines, analytical estimates~\cite{noteSUPPL} with an overall fitting multiplicative factor. 
In all panels: averages over 50 realizations, ${\cal B}=0$,  $ \alpha=0$. 
}
\label{fig:Pleak}
\end{figure}

In Figs.~\ref{fig:Pleak} (a) and (b), we  compute the probability, $P_b(t)$, for the initial state to remain in its original energy band $b$,
\begin{equation}
P_b(t) = \sum_k |\langle V_k^b |e^{-i H t} | \Psi(0)\rangle|^2,
\end{equation}
where the sum includes all the states of the selected energy band. The results are shown for $\langle P_b(t) \rangle$, where $\langle . \rangle$ indicates average over random realizations and initial states. We show the case of $b=1$, but similar results hold for other bands.
It is evident that the probability to remain in the initial band increases with system size.
This happens in the presence of a random transverse field [Fig.~\ref{fig:Pleak} (a)]  and also when NN interactions are added  [Fig.~\ref{fig:Pleak} (b)]. 

In Figs.~\ref{fig:Pleak} (c) and (d), we plot the asymptotic values of the leakage probability, $P_{leak}=1- \lim_{t\to\infty}\langle P_b(t)\rangle$, as a function of the random field strength for $J_z=0$ [Fig.~\ref{fig:Pleak} (c)] and {\em vs} the NN coupling strength for $W=0$ [Fig.~\ref{fig:Pleak} (d)]. $P_{leak}$ represents the probability for $| \Psi(0)\rangle$ to leak outside its original band. It decreases with $L$, showing that as the system size increases, the evolution of $|\Psi(0)\rangle$ remains more and more confined to a subspace of $V$ for a longer time. Note that the distance between the bands nearby the initial one increases with $L$, but so does the number of states which are connected by $H_0$. The suppression of leakage takes into account this  non-trivial interplay.
A perturbative argument leads 
to $P_{leak} \propto (W/J)^2/L$ for $W\ne 0$ and $J_z=0$, 
while $P_{leak} \propto (J_z/J)^2/L$ for  NN interaction only~\cite{noteSUPPL}.
Such scaling relations are consistent with our numerical data in Figs.~\ref{fig:Pleak}  (c) and (d). 

The invariant subspaces generated by long-range interaction can be related to the QZE~\cite{Peres1980,Facchi2001,Facchi2001b,Facchi2002,FacchiProceed}. This term refers to the familiar freezing of the dynamics due to frequent measurements, but also to the onset of invariant Zeno subspaces that occurs in unitary dynamics due to strong interactions~\cite{Facchi2001b,FacchiProceed} and which has been studied experimentally~\cite{Raimond2012}. The latter is closer to our case and can be explained as follows. Consider the total Hamiltonian $H=H_s +gH_{\rm meas}$, which one may interpret as a quantum system described by $H_s$ that is {\em continuously} observed by an ``apparatus'' characterized by $gH_{\rm meas}$. In the limit of strong coupling, $g\rightarrow \infty$, a superselection rule is induced that splits the Hilbert space into the eigensubspaces of $H_{\rm meas}$. Each one of these invariant quantum Zeno subspaces is specified by an eigenvalue $v_k$  and is formed by the corresponding set of degenerate eigenstates of $H_{\rm meas}$. The dynamics becomes confined to these subspaces and dictated by the Zeno Hamiltonian $H_Z = \sum_k \Pi_k H_s \Pi_k + v_k \Pi_k$, where $\Pi_k$ are the projectors onto the eigensubspaces of $H_{\rm meas}$ corresponding to the eigenvalues $v_k$. 

For the system investigated here, we associate $H_s$ with $H_0$ and $gH_{\rm meas}$ with $V$. 
The subspaces of $V$, with fixed numbers $b$ of excitations,  become invariant subspaces of the total Hamiltonian not only when $J\to\infty$ with ${\cal B}, W, J_z$ fixed, which is the scenario of the QZE described above, but also in the large system size limit, $L \rightarrow \infty$, which is the main focus of this work. 

When $J_z=0$, the Zeno Hamiltonian coincides with $V$,  because the transverse field does not couple directly states $|V^b_k\rangle$ that belong to the same eigensubspaces of $V$, so $\sum_k \Pi_k H_0 \Pi_k = 0$.  This explains why the dynamics in Fig.~\ref{fig:cone} (b) is frozen for very long times. On the other hand, in the case where ${\cal B}, W=0$ and $J_z \ne 0$, we can rewrite $H_0$ in terms of the $\sigma^{\pm_x}_n$ operators that flip the spins in the $x$-direction. The projection of the NN part of the Hamiltonian on the eigensubspaces of $V$ leaves only the term $\sigma_{n}^{+_x} \sigma_{n+1}^{-_x}+\sigma_{n}^{-_x} \sigma_{n+1}^{+_x}$, which leads to a Zeno Hamiltonian with an effective NN interaction that conserves the number of excitations inside each band $b$. This explains why in Fig.~\ref{fig:cone} (d) a light cone typical of short-range interactions appears. 

\begin{figure}[htb]
\centering
\includegraphics*[width=3.0in]{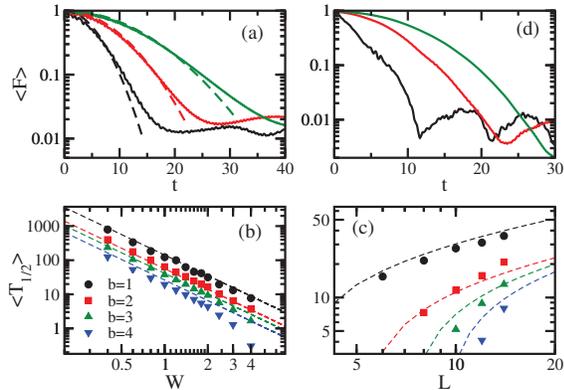}
\caption{(Color online) Fidelity decay and time for it to reach the value 1/2; initial states are random superpositions of $| V_k^b \rangle$. Upper panels: $F(t)$ for $b=3$ for $J_z=0, W=2$ (a) and for $W=0, J_z=1$ (d). From bottom to top: $L=10,12,14$. Numerical results: full lines. Gaussian decay: dashed lines. Lower panels have $J_z=0$ and give  $T_{1/2}$ {\it vs} $W$ for $L=12$ (b), and {\it vs} $L$ for $W=2$ (c), for $|\Psi(0)\rangle$ from different bands.
Numerical data: symbols. Analytical estimate $T_{1/2} = c_1 /\delta E$  with $c_1$ a fitting parameter: dashed lines. All panels: averages over 50 realizations, $\alpha=0$, ${\cal B}=0$.
}
\label{fig:Fid}
\end{figure}

{\em Fidelity Decay.--} 
To substantiate that the dynamics in the subspaces with fixed $b$ becomes indeed controlled by the Zeno Hamiltonian as $L$ increases, we analyze the fidelity between an initial state evolved under the total Hamiltonian $H$ and the same state evolved under $H_Z$,
\begin{equation}
F(t) = |\langle \Psi(0) |e^{i H_Z t} e^{-i H t} | \Psi(0)\rangle|^2.
\label{eq:Fid}
\end{equation}
It is clear that if $H\rightarrow H_Z$ then $F(t) \rightarrow 1$. 
The results are shown in Fig.~\ref{fig:Fid}.
Equivalently to  Fig.~\ref{fig:Pleak}, we fix ${\cal B}=0$ and deal with averages over disorder and initial states, which gives $\langle F(t) \rangle$. $| \Psi(0)\rangle$ is again a random superposition of all states $|V_k^b\rangle$ belonging to the same band $b$. 

In Figs.~\ref{fig:Fid} (a) and (d) the fidelity is plotted {\it vs} time for different system sizes for the band with  $b=3$. In panel (a), $H_0$ contains only the random fields, while in (d), $H_0$ contains only NN interaction. In both cases the fidelity decay slows down as the system size increases, confirming that $H_Z$ determines the dynamics for large $L$. 

For the  $J_z=0$ case of Fig.~\ref{fig:Fid} (a), since the projection of $H_0$ on the $b$ subspace is zero, the fidelity coincides with the survival probability, $F(t) = |\langle \Psi(0) | \Psi(t)\rangle|^2 $, which, counterintuitively, decays slower as the system size increases. This shows that the dynamics localizes as $L\rightarrow \infty$. $F(t)$ decays as a Gaussian~\cite{Izrailev2006,TorresAll2014b,TorresAll2014b2,TorresAll2014b3,TorresAll2014b4} -- see dashed lines in Fig.~\ref{fig:Fid} (a).

In Figs.~\ref{fig:Fid} (b) and (c) we study how the time $T_{1/2}$ that it takes for the survival probability to reach the value $1/2$ 
depends on the disorder strength (b) and on system size (c). Figure~\ref{fig:Fid} (b) provides information associated with the usual QZE, where the quantum Zeno subspaces are induced by decreasing the strength of $H_0$. One sees that the dynamics slows down with the reduction of disorder as $\langle T_{1/2} \rangle \propto W^{-2}$. In Fig~\ref{fig:Fid} (c), $\langle T_{1/2}\rangle$ grows with $L$, corroborating our claims that the fidelity increases and the excitations become more localized as the system size increases. 

The estimation of the dependence of $T_{1/2}$ on the parameters of $H$ goes as follows.  Since
the eigenstates of $V$ in each invariant subspace are degenerate,  the perturbation $H_0$ mixes them
all. In this case, the energy uncertainty $\omega$ of the initial state can be approximated by the energy spread 
$\delta E$ of each band induced by the perturbation. 
The fidelity decay can then be estimated as $T_{1/2} \simeq 1/\delta E$, where $\delta E$ is computed from 
perturbation theory~\cite{noteSUPPL}. For large system sizes one has $T_{1/2} \propto J\sqrt{L}/W^2$. The analytical estimates for $T_{1/2}$ are shown with dashed curves in Figs.~\ref{fig:Fid} (b) and (c). The agreement is excellent.

We note that $T_{1/2}$ gives the time scale over which the shielding effect persists. In finite systems,  shielding is effective for a finite time that can, however, be exceedingly long, as shown in Fig.~\ref{fig:Fid}.

{\em Conclusions.--}  
We revealed a generic effect of long-range interacting systems: {\it Cooperative Shielding}. It refers to invariant subspaces that emerge as the system size increases. Inside these subspaces, the dynamics occurs as if long-range interaction was absent, being dictated by effective short-ranged Hamiltonians. A parallel was established between these Hamiltonians and  Zeno Hamiltonians.

The analysis and control of nonequilibrium dynamics can never be detached from the initial state considered. For exactly the same Hamiltonian with long-range interaction, an initial state with components in the various subspaces induced by that interaction leads to a nonlocal propagation of perturbation, as demonstrated experimentally with ion traps~\cite{Jurcevi2014,Richerme2014}, while an initial state belonging to a single subspace is unaffected by the long-range interaction, as verified here. {\it Cooperative Shielding} could also be tested by those experiments. 

\begin{acknowledgments}
We acknowledge useful discussions with R. Bachelard, A. Biella, G. G. Giusteri, F. Izrailev, R. Kaiser, S. Ruffo and R. Trasarti-Battistoni. This work was supported by the  NSF grant No.~DMR-1147430. 
\end{acknowledgments}


\onecolumngrid
\vspace*{0.4cm}
\begin{center}
{\large \bf Supplementary material for EPAPS} 
\vskip 0.2 cm
{\large \bf Cooperative shielding in many-body systems with  long-range interaction} \\
\vspace{0.6cm}
Lea F. Santos$^1$, Fausto Borgonovi$^2$, and Giuseppe Luca Celardo$^2$\\
$^1${\it Department of Physics, Yeshiva University, New York, New York 10016, USA \\
ITAMP, Harvard-Smithsonian Center for Astrophysics, Cambridge, MA 02138, USA} \\
$^2${\it Dipartimento di Matematica e Fisica and ILAMP, 
Universit\'a Cattolica del Sacro Cuore, Brescia, ITALY \\
Istituto Nazionale di Fisica Nucleare, sez. Pavia, Pavia, ITALY}

\end{center}
\vspace{0.6cm}
\twocolumngrid

\section{Introduction to the Supplementary Material}

Here, we provide further illustrations, reinforcing that shielding is a generic property of long-range interacting systems. We also analyze the magnetic reversal  of the spins in the presence of long-range interactions and show how we obtained our estimates for the leakage probability and the energy spread $\delta E$ of each band. The energy spread is used for approximating $T_{1/2} \simeq 1/\delta E$, which refers to time for the survival probability to reach the value 1/2.

\section{Shielding Effect}
\begin{figure}[ht!]
\centering
\includegraphics*[width=3.in]{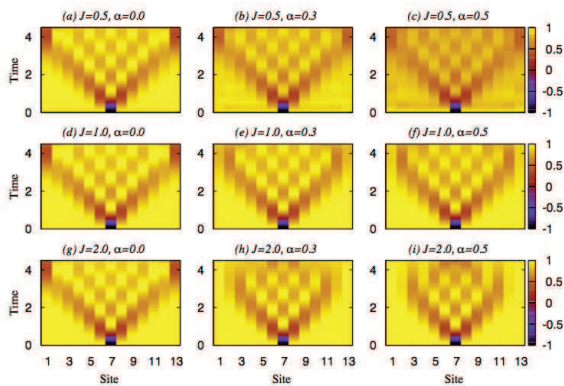}
\caption{(Color online) Evolution of the polarization $\langle \sigma_n^x (t) \rangle$ for all sites $n$; $L=13$, ${\cal B}=1/2$, $W=0$, and $J_z=1$. $J$ increases from top to bottom and $\alpha$ from left to right. Initial state: $\langle \sigma_{(L+1)/2}^x (0)\rangle =-1 $ and $\langle \sigma_{n \neq (L+1)/2}^x (0) \rangle = + 1$.}
\label{fig:alpha}
\end{figure}

To further support that in the presence of long-range interaction $(\alpha<1)$, the dynamics is shielded from $V$ and therefore does not depend on the long-range coupling strength $J$ for a long time, we show in Fig.~\ref{fig:alpha} the evolution of the spin polarization in the $x$-direction for different values of $J$. The initial state has the spins aligned in the $x$-direction as in the Fig.1 of the main text. We fix the strength of the NN interaction to $J_z=1$. The figure shows that the speed of the propagation remains unchanged, as $J$ increases from top to bottom. It depends only on the strength $J_z$ of the NN coupling. The same behavior is observed also as   $\alpha$ increases from left to right.

\section{Magnetic Reversal}
\begin{figure}[ht!]
\centering
\includegraphics*[width=3.in]{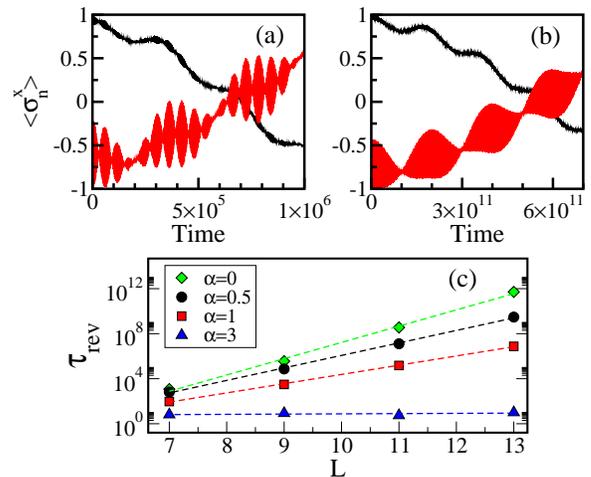}
\caption{(Color online) Evolution of the polarization $\langle \sigma_n^x (t) \rangle$ for all sites $n$ (top) and reversal time $\tau_{rev}$ of the central spin as a function of the system size, $L$ for different interaction ranges (bottom). Fized parameters: ${\cal B}=1/2$; $W=0$, $J_z=0$. Initial state: $\langle \sigma_{(L+1)/2}^x (0)\rangle =-1 $ and $\langle \sigma_{n \neq (L+1)/2}^x (0) \rangle = + 1$. Panels (a) and (b): $\alpha=0$; $L=9$ for (a) and $L=13$ for (b). Dark (black) curve: all sites $n \neq (L+1)/2$ and light (red) curve: $n =(L+1)/2$. The sign-reversal of the polarization occurs at the same time for all spins. }
\label{fig:sync}
\end{figure}

The frozen dynamics seen in Figs.1 (b) and (c ) of the main text, for ${\cal B}\neq0$ and $J_z=1$, holds for a finite time. Eventually the spins reverse their signs.  The polarized-sign reversal for $\alpha<1$ is a collective effect: the external field rotates all spins synchronously in the $xy$-plane, as seen in Figs.~\ref{fig:sync} (a) and (b) below.

In Fig.~\ref{fig:sync} (c),  we analyze the time $\tau_{rev}$ for the polarization of the central site to change sign. A strong dependence on the interaction range is found: $\tau_{rev}$ increases exponentially with $L$ for $\alpha \le 1$, while  for short-range interaction, it is independent of system size. The exponential dependence of the reversal time
on the system size is a consequence of the fact that long-range interactions
induce a polynomially large energy barrier, which must be overcome for the magnetization
to reverse its sign. Such barrier induces ergodicity breaking in classical systems
with long-range interactions (see Ref.[18] of the main text). On the quantum side, such
barrier can be overcome through the macroscopic quantum tunneling of the magnetization
(see Ref.[17] of the main text). Since the energy barrier is polynomial in the system size,
an exponential tunneling rate follows, which explains the results presented in Fig.~\ref{fig:sync} (c).  Notice that $\tau_{rev}$ is distinct from the time $T_{1/2}$ for the fidelity to reach the value 1/2. The latter grows polynomially with system size.

\section{Leakage Probability}

$P_{leak}$  refers to the probability for an initial state
$|V\rangle$, corresponding to an eigenstate of $V$,  [see Eq.~(1) in
the main text] to leak outside its original energy band. The leakage probability is defined as 
$P_{leak}=1-\lim_{t\to\infty} P_b(t)$, where $P_b$ is the probability for the initial state to remain in the band $b$ [see Eq.~(3) in the main text].

If we start from a certain quantum mechanical state coupled with an
amplitude $\epsilon$  to another quantum mechanical state, the two being separated by
an energy $\Delta$, the probability to find the system on the second
state will never be one; instead, for $\epsilon/\Delta \ll 1$, it will be at most of the order
$(\epsilon/\Delta)^2$. Having this in mind we can estimate the asymptotic value of the leakage probability. 
The results for our estimates are valid in the dilute limit when the
number of excitations is small compared to the system size, $b/L \ll
1$.

\subsection{In the presence of an external field and no NN coupling}

For the Hamiltonian given by the Eq.~(2) of the main text, with an external field and no NN coupling ($J_z=0$), each state $|V\rangle$  in
a band $b$ is connected to approximatively $L$ states in the
nearest-neighboring bands. The coupling amplitude is $h_n$, with $\langle
h_n^2 \rangle= W^2/12$. Thus, we can define the strength of the
coupling as,
$$
\epsilon \simeq \frac{W}{\sqrt{12}} .
$$
For small $b$'s, the energy distance between the neighboring bands is proportional to $J L$ (see main text), therefore
\begin{equation}
P_{leak} \propto L \frac{\epsilon^2}{\Delta^2} \propto \frac{W^2}{J^2 L}.
\label{PleakApprox}
\end{equation}

As an example, let us consider an initial state $|V\rangle$ from band 
$b=1$, where all spins point up in the $x$-direction, except for one spin, which points down.  The term of the Hamiltonian $H_0^W$ containing the external field can be written as,
\begin{equation}
H_0^W =  \sum_{n=1}^L h_n  \sigma_n^z  = \sum_{n=1}^L  \frac{h_n}{2}  ( \sigma_{n}^{+,x} + \sigma_{n}^{-,x}),
\label{eq:Wsigma}
\end{equation}
where the operators $\sigma_n^{{\pm},x}$ flip the spins pointing in the $x$-direction.
Due to this term, our initial state  is connected with one state in
band $b=0$ and with $L-1$ states in band $b=2$, so that we have: 
$$
P_{leak}^{b=1} \simeq \frac{\epsilon^2}{\Delta_{0,1}^2} + (L-1) \frac{\epsilon^2}{\Delta_{1,2}^2} ,
$$
where  $\Delta_{0,1}= 2J (L-1)$ and $\Delta_{1,2}= 2J(L-3)$ are the energy differences between band $b=1$ and the two neighboring bands.

In general, starting with an initial state  $|V\rangle$ in a band $b$, there are $b$ connections
with the $b-1$ band and $L-b$ connections with the $b+1$ band, so that 
\begin{equation}
P_{leak}^b \simeq  b \frac{\epsilon^2}{\Delta_{b-1.b}^2} + (L-b) \frac{\epsilon^2}{\Delta_{b,b+1}^2} ,
\label{pw}
\end{equation}
where $\Delta_{b-1,b}  = E_{b-1} - E_b = 2J (L-2b+1)$. 
This expression confirms the general scaling of $P_{leak}$ given by
Eq.~(\ref{PleakApprox}).

\subsection{In the presence of NN coupling and no external field}

We can estimate $P_{leak}$ also for  $J_z \ne 0$. 
The term of $H_0$ containing the NN interaction can be written as,
\begin{equation}
H_0^{J_z}=\sum_{n=1}^{L-1} J_z \sigma_n^z  \sigma_{n+1}^z = \sum_{n=1}^{L-1} \frac{ J_z}{4}  (\sigma_n^{+,x}+ \sigma_n^{-,x})(\sigma_{n+1}^{+,x}+ \sigma_{n+1}^{-,x} ).
\label{eq:Jzsigma}
\end{equation}
In general, $J_z$ connects a state in band $b$ with $m$ states inside that same band,
$n^+$ states in band $b+2$, and $n^-$ states in band $b-2$, such that
$m+n^+ +n^- \simeq L$. For a typical state with $b$ separated
excitations, each one placed at 
least one site apart from the other,
there are $2b$ states connected in the same band and
$L-2b-1$ states connected in the outer bands, so that for $b/L \ll 1$
the ratio $n^{\pm }/m \propto L$. Since the amplitude of the coupling
is $J_z$, we can write 
$$
P^b_{leak} = \frac{n_{+}}{m+1} \frac{J_z^2}{\Delta_{b+2,b}^2} +
\frac{n_{-}}{m+1} \frac{J_z^2}{\Delta_{b-2,b}^2} \propto
\frac{J_z^2}{J^2 L} 
$$

Let us compute $P_{leak}$ explicitly for the band $b=1$. 
An initial state $|V\rangle$ from band $b=1$ is coupled to two other states inside that band (apart from the situation where the initial state has an excitation on one of the border sites, 1 or $L$, in which case it couples only with one other state) and $L-3$ states in band $b=3$ (for an excitation on
the border we have $L-2$ connections), so that we have: 
\begin{equation}
P_{leak}^{b=1} \simeq \frac{(L-3)}{2} \frac{J_z^2}{\Delta_{3,1}^2}= \frac{(L-3)}{8}
\frac{J_z^2}{J^2 (2L-8)^2} .
\label{pnn}
\end{equation}

\section{Energy spread}

The time that it takes for the survival probability to reach the value
1/2 is denoted by $T_{1/2}$. As discussed in the main text, the
survival probability decay shows a Gaussian behavior, which justifies
writing $T_{1/2} \sim 1/\omega$, where $\omega$ is the energy
uncertainty of the initial state.  Since the eigenstates of $V$ in each band are degenerate,  the perturbation $H_0$ mixes them
all. In this case, the energy uncertainty $\omega$ of the initial state can be approximated by the energy spread $\delta E$ in each band induced by the 
perturbation. The analytical expression for the  energy spread $\delta E $ 
evaluated at the first nonzero order of perturbation theory is studied below.

We compute the eigenvalues of the total Hamiltonian given by Eq.~(2) of the main text, 
using second order perturbation theory for degenerate levels. 
Note that our unperturbed Hamiltonian is the long-range part $V$, 
while we consider $H_0$ as the perturbation. 
In the following, we set ${\cal B},J_z=0$, so that the perturbation $H_0$ is determined only by the random field, as in Eq.~(\ref{eq:Wsigma}) above.

Let us consider the energy band $b=1$.
We denote the initial state $|V\rangle$ in this band as $|1,k\rangle$, where
$k=1,...,L$ indicates the position of the excitation. 
In labeling the states, we neglect
the double degeneracy due to the flipping of all  spins, since the states with $M_x=(L-2)/2$ are only connected to those with $M_x=-(L-2)/2$ in a very high order of perturbation theory.

The perturbation $H_0^W$ [Eq.~(\ref{eq:Wsigma}) above]  connects the initial state in band $b=1$ to the state $|0,0\rangle$, belonging to  band $b=0$, and to $L-1$ states in 
band $b=2$. The latter states are denoted by $|2,k,j\rangle$; they have one excitation on site $k=1,...,L-1$ and the other on site
$j=k+1,...,L$, so that the total number of states is
$L(L-1)/2$.

Since the degeneracy inside the band
is not removed at first order of perturbation theory,
we use second order perturbation theory for degenerate levels, namely
the eigenvalue problem,
\begin{equation}
\label{eq:sm1}
(V+ \epsilon H_0^W) |\psi_1\rangle = (E_1 + \epsilon E^I_1+
\epsilon^2 E^{II}_1) |\psi_1\rangle
\end{equation}
where
\begin{eqnarray}
|\psi_1\rangle &&=  \sum_{k=1}^{L} c_{0,k} |1,k\rangle
+(\epsilon  c_{-,0}^I +\epsilon^2  c_{-,0}^{II}) |0,0\rangle
\label{eq:sm2}
\\
&&+ 
\sum_{k=1}^{L-1}  \sum_{j=k+1}^L ( \epsilon c_{+,k,j}^I 
+\epsilon^2  c_{+,k,j}^{II}) |2,k,j\rangle\nonumber.
\end{eqnarray}
The action of the ``unperturbed'' Hamiltonian on the unperturbed states is
trivial,
$$
V |b,k\rangle = E_b |b,k\rangle, \quad {\rm with} \quad b=0,1,2,...
$$
while  the perturbation $H_0^W$ acts as,
\begin{equation}
\label{eq:sm3}
 H_0^W | 1, k \rangle = h_k |0,0\rangle
+  \sum_{j\ne k} h_j |2,k,j\rangle.
\end{equation}
Collecting  the $\epsilon$ terms in Eq.~(\ref{eq:sm1}) gives,
\begin{eqnarray}
\label{eq:sm4}
&& - E_1^I \sum_{k=1}^{L} c_{0,k} |1,k\rangle+
(E_{0}-E_1) c_{-,0}^I |0,0\rangle +\nonumber\\ 
&&(E_{2}-E_1) \sum_{k=1}^{L-1}  \sum_{j=k+1}^L  c_{+,k,j}^I |2,k,j \rangle+ \nonumber \\
&&\sum_{k=1}^{L} c_{0,k} \left( 
h_k |0,0\rangle
+  \sum_{j\ne k} h_j |2,k,j\rangle \right) =0
\end{eqnarray}
Bracketing Eq.~(\ref{eq:sm4}) respectively  with 
$ \langle  0,0|$,  $ \langle  1,s|$ and $ \langle 
 2,\alpha,\beta|$ with $\beta > \alpha$, we
obtain,
\begin{eqnarray}
&&c_{-,0}^I = \frac{1}{E_1-E_0}\sum_{k=1}^{L} c_{0,k} 
h_k, \label{eq:sm5a}\\
&&E_1^I = 0,  \nonumber\\
&& c_{+,\alpha,\beta}^I = \frac{1}{E_1-E_2}\left(
c_{0,\alpha}h_\beta+
c_{0,\beta} h_\alpha\right).\label{eq:sm5b}
\end{eqnarray} 
Collecting  the $\epsilon^2$ terms in Eq.~(\ref{eq:sm1}) and taking into account that $E_1^I = 0$, we have,
\begin{eqnarray}
\label{eq:sm6}
&&(E_{0}-E_1) c_{-,0}^{II} |0,0\rangle -
E_1^{II} \sum_{k=1}^{L} c_{0,k} |1,k\rangle+\nonumber\\
&&(E_{2}-E_1) \sum_{k=1}^{L-1}  \sum_{j=k+1}^L  c_{+,k,j}^{II} 
|2,k,j \rangle+ \nonumber \\
&& c_{-,0}^I H_0^W |0,0\rangle
+ \sum_{k=1}^{L-1}  \sum_{j=k+1}^L c_{+,k,j}^I  H_0^W|2,k,j\rangle
=0.\nonumber\\ 
&& 
\end{eqnarray}
Bracketing Eq.~(\ref{eq:sm6}) with $ \langle  0,0|$,  $ \langle  1,s|$ 
and $ \langle  2,\alpha,\beta|$ with $\beta > \alpha$, we get $c_{-,0}^{II} = 0$, $c_{+,\alpha,\beta}^{II} = 0$, and
\begin{equation}
E_1^{II} c_{0,s} =  c_{-,0}^I \langle 1,s |H_0^W |0,0\rangle+
\sum_{j>k} c_{+,k,j}^I \langle 1,s | H_0^W|2,k,j\rangle,
\label{eq:sm7}
\end{equation}
The equation above, due to the symmetry
of the coefficients $c_{+,k,j}^I= c_{+,j,k}^I $ [see Eq.~(\ref{eq:sm5b})]
 can be rewritten as, 
\begin{equation}
\label{eq:sm8}
E_1^{II} c_{0,s} = c_{-,0}^I h_s + \sum_{j\ne s} c_{+,s,j}^I h_j.
\end{equation}
Inserting  Eq.~(\ref{eq:sm5a}) and Eq.~(\ref{eq:sm5b}) in Eq.~(\ref{eq:sm8}), one
finds that the second order corrections $E_1^{II}$
are the solutions of the $L$ equations,
\begin{eqnarray}
&&c_{0,s} \left[ \frac{h^2_s}{(E_1-E_0)} + \sum_{k\ne s} \frac{h^2_k}{(E_1-E_2)}- E_1^{II} \right]+\nonumber\\
&& \sum_{j\ne s} c_{0,j} h_s h_j\left[ \frac{1}{(E_1-E_0)} +  \frac{1}{(E_1-E_2)}\right] =0,\nonumber\\
&&
\end{eqnarray}
with $s=1,..,L$. In other words, 
$E_1^{II}$ are the eigenvalues of the symmetric matrix $C$, whose diagonal elements are,
\begin{equation}
C_{ss} = \frac{h^2_s}{(E_1-E_0)} + \sum_{k\ne s} \frac{h^2_k}{(E_1-E_2)},
\label{sm:diag}
\end{equation}
and off-diagonal elements for $k\neq s$ are, 
\begin{equation}
C_{ks} =  h_k h_s\left[ \frac{1}{(E_1-E_0)} +  \frac{1}{(E_1-E_2)}\right].
\label{sm:off}
\end{equation}

Let us now estimate the eigenvalues of this matrix in the limit of large $L$,
at fixed $W$. We know that,
$$
E_1-E_0 = 2J(1-L) \qquad {\rm and} \quad E_1-E_2= 2J(L-3),
$$
therefore, in the limit of large $L$ one has 
that the off-diagonal elements are negligible with 
respect to the diagonal ones, since
$C_{ks} \sim o(1/L)$ for $k\ne s$, while $C_{ss} \sim o(1)$.

Thus, we can estimate
the eigenvalues from the diagonal elements only. In particular, since we
are interested in the energy spreading, we  evaluate
\begin{eqnarray}
&&\delta E^2  =   \langle C_{ss}^2 \rangle_{_W} - \langle C_{ss} \rangle_{_W}^2 \nonumber \\
&&= \frac{W^4}{180 J^2} \left( \frac{1}{(L-1)^2}+\frac{L-1}
{(L-3)^2} \right).
\end{eqnarray}
where in computing the average over disorder we took into account that
$$
\langle h_s h_k \rangle_{_W} = 0 \quad {\rm for }  \quad s\ne k,
$$
and
$$
\langle h_s^2 \rangle_{_W} = \frac{W^2}{12},
\qquad  
\langle h_s^4 \rangle_{_W} = \frac{W^4}{80}.
$$

In the limit of large system size, we therefore get the asymptotic
behavior,
\begin{equation}
\delta E \sim \frac{W^2}{J \sqrt{L}} .
\label{estimat}
\end{equation}

The generalization to an arbitrary band $b$ is far from trivial. We will provide the details of this derivation in a longer version of this paper.  We present here only the final result. 
Similar to the case $b=1$, we can estimate the energy spreading
for the general $b$ band as,
\begin{eqnarray}\label{eq:ge1}
\delta E^2    
= \frac{W^4}{180 J^2} \left( \frac{b}{(2b-L-1)^2}+\frac{L-b}
{(L-2b-1)^2} \right).
\end{eqnarray}
which gives the same estimate as in Eq.~(\ref{estimat}).

\end{document}